\documentclass{sig-alternate}

\DeclareMathOperator*{\argmax}{arg\,max}

\usepackage{ctable}
\usepackage{multirow}
\usepackage{balance}
\usepackage{booktabs}
\usepackage{hyphsubst}
\usepackage{graphicx}
\usepackage{amsmath}
\usepackage{url}
\usepackage{hhline}
\usepackage[caption = false]{subfig}

\permission{Copyright is held by the author.}

\begin{document}

\title{Modeling Cognitive Processes in Social Tagging to Improve Tag Recommendations}

\numberofauthors{1}
\author{
Dominik Kowald\\
       \affaddr{Supervised by: Prof. Stefanie Lindstaedt (Know-Center)}\\
       \affaddr{Know-Center, Graz University of Technology}\\
       \affaddr{Inffeldgasse 13, Graz, Austria}\\
       \email{dkowald@know-center.at}
}

\maketitle
\begin{abstract}
With the emergence of Web 2.0, tag recommenders have become important tools, which aim to support users in finding descriptive tags for their bookmarked resources. Although current algorithms provide good results in terms of tag prediction accuracy, they are often designed in a data-driven way and thus, lack a thorough understanding of the cognitive processes that play a role when people assign tags to resources. This thesis aims at modeling these cognitive dynamics in social tagging in order to improve tag recommendations and to better understand the underlying processes.

As a first attempt in this direction, we have implemented an interplay between individual micro-level (e.g., categorizing resources or temporal dynamics) and collective macro-level (e.g., imitating other users' tags) processes in the form of a novel tag recommender algorithm. The preliminary results for datasets gathered from BibSonomy, CiteULike and Delicious show that our proposed approach can outperform current state-of-the-art algorithms, such as Collaborative Filtering, FolkRank or Pairwise Interaction Tensor Factorization. We conclude that recommender systems can be improved by incorporating related principles of human cognition.
\end{abstract}



\section{Introduction} \label{sec:intro}
Social tagging systems enable users to collaboratively assign freely chosen keywords, so-called tags, to resources. These tags can then be used for navigating, searching, organizing and finding content, and serendipitous browsing \cite{jaschke2008tag,Korner2010}. Hence, tags have become an essential instrument of Web 2.0, the social Web, assisting users during these activities. While in social tagging systems users can freely choose keywords for their bookmarked resources, they have to create a set of descriptive tags on their own, which can be a demanding task \cite{lipczak2012hybrid}.

As a solution, a variety of tag recommender algorithms, such as Collaborative Filtering, FolkRank or Pairwise Interaction Tensor Factorization, have been proposed. Tag recommenders suggest a set of tags for a given user-resource pair based on previously used and assigned tags and aim at helping not only the individual to find appropriate tags but also the collective to consolidate the shared tag vocabulary \cite{lipczak2012hybrid}. Furthermore, Dellschaft \& Staab \cite{Dellschaft2012} have shown that personalized tag recommenders can increase the indexing quality of resources, making it easier for users to understand the information content of an indexed resource based on its assigned tags.

\subsection{Problem Statement} \label{sec:problems}
Although current state-of-the-art tag recommender approaches (see Section \ref{sec:relwork}) perform reasonably well in terms of recommender accuracy, most of them are designed purely data-driven. As a result, they are based on either simply counting tag frequencies or computationally expensive calculation steps (e.g., calculating user similarities or factorizing entities). Hence, these approaches typically ignore important insights originating from cognitive research of how people assign words or tags to resources, which is essential for the design of tag recommenders, that should attempt to mimic the user's tagging behavior.

As a prominent example in this respect, the work proposed by Fu \cite{fu2008microstructures} discusses an interplay between individual micro-level (e.g., categorizing resources or temporal dynamics) and collective macro-level (e.g., imitating other users' tags) processes in social tagging systems (see Section \ref{sec:approach}). Based on that, we state the hypothesis, that a theory-driven approach that is build upon such insights can not only improve recommender accuracy in general but also can help to better understand the underlying cognitive processes.

\subsection{Contributions} \label{sec:cont}
This thesis aims to develop a novel tag recommender approach, which models the cognitive processes that play a role when people assign tags to resources. At the current stage, our contributions are as follows:
\begin{itemize}
	\item We propose a novel theory-driven approach for recommending tags, which models cognitive processes in social tagging in order to mimic the way humans assign tags to resources.
	\item We conduct an extensive evaluation using dataset samples gathered from three real-world folksonomies (BibSonomy, CiteULike and Delicious) to show the effectiveness of our theory-driven approach.
	\item We show that our approach can outperform several state-of-the-art tag recommender algorithms, such as FolkRank or Pairwise Interaction Tensor Factorization, in terms of recommender accuracy.
	\item We introduce an open-source tag recommender benchmarking framework termed TagRec, which contains not only our proposed approach but also standardized baseline algorithms and evaluation methods.
\end{itemize}

\section{Related Work} \label{sec:relwork}
To date, two types of tag recommenders have been established: folksonomy- and content-based approaches \cite{lipczak2012hybrid}. At the moment, in this work we focus on folksonomy-based algorithms. The most basic approach in this respect is the unpersonalized \textit{MostPopular (MP)} algorithm that recommends for any user and any resource the same set of tags weighted by the frequency in all tag assignments \cite{jaschke2007tag}. A personalized extension of MP is the  \textit{MostPopular$_{u, r}$ (MP$_{u,r}$)} algorithm that suggests the most frequent tags in the tag assignments of the user, \textit{MostPopular$_{u}$ (MP$_{u}$)}, and the resource, \textit{MostPopular$_{r}$ (MP$_{r}$)} \cite{jaschke2007tag}. Another classic recommender approach is \textit{Collaborative Filtering (CF)}, which has been adapted for tag recommendations by Marinho et al. \cite{marinho2008collaborative} to form the neighborhood of a user based on the tag assignments in the user profile. According to Gemmell et al. \cite{gemmell2009improving}, the best results for CF in social tagging systems are obtained with a neighborhood size of 20 users.

Another well-known tag recommender approach is the \textit{FolkRank (FR)} algorithm, which is an improvement of the \textit{Adapted PageRank (APR)} approach \cite{hotho2006information}. FR and APR adapt the Google PageRank algorithm to rank the nodes within the graph structure of a folksonomy based on their importance in the network \cite{jaschke2007tag}. A different popular and recent tag recommender mechanism is \textit{Pairwise Interaction Tensor Factorization (PITF)} proposed by Rendle \& Schmidt-Thieme \cite{Rendle2010}, which is an extension of \textit{Factorization Machines (FM)} \cite{rendle2010factorization} and explicitly models the pairwise interactions between users, resources and tags in order to predict future tag assignments. As for algorithms that utilize topic models, to date mainly methods have been proposed based on \textit{Latent Dirichlet Allocation (LDA)} (e.g., \cite{krestel2009latent}).

With regard to time-dependent tag recommenders, there are two notable approaches. First, the \textit{GIRPTM} algorithm presented by Zhang et al. \cite{zhang2012integrating}, which considers the frequency and the temporal usage of a user's tag assignments. GIRPTM models the temporal tag usage via an exponential distribution. Second, the \textit{BLL+C} algorithm introduced in \cite{kowald2015refining}, which incorporates the activation equation from the cognitive model ACT-R proposed by Anderson et al. \cite{Anderson2004} and uses a power-law function to mimic the temporal decay of tag reuse. Both approaches imitate tagging by simply taking into account the most popular tags previously assigned to the resource (MP$_r$). Kowald et al. \cite{kowald2015refining} have demonstrated that BLL+C outperforms GIRPTM and other well-established algorithms, such as CF, FR and PITF.

\section{Proposed Approach} \label{sec:approach}
Our approach is based on an interplay between micro-level (i.e., the individual level) and macro-level (i.e., the collective level) processes in social tagging systems (e.g., \cite{fu2008microstructures}). Among others, micro-level processes (see Sections \ref{sec:categorize} and \ref{sec:temporal}) involve categorizing a resource (e.g., modeled as LDA topics) and turning the latent (i.e., non-observable) categorization into manifest (i.e., observable) words or tags \cite{kintsch2011construction}. Beyond that, temporal dynamics have turned out to influence the choice of words for a given resource, i.e., recently used tags have a higher probability of being reused than ``older'' ones \cite{polyn2009context}. The effect of macro-level structures (see Section \ref{sec:imitate}) is mediated by the user's tendency to imitate other users' tags \cite{floeck2011imitation,Seitlinger2012}.

In this section we describe how we modeled and implemented these micro-macro dynamics and the corresponding cognitive processes in the form of a novel tag recommender approach. Needless to say, there are also other types of dynamics and processes (e.g., decision making or associative memory activations \cite{Anderson2004}) that play a role when people choose tags for resources but at the current stage of this thesis, we focus on the mentioned ones.

\begin{figure}[t!]
        \centering
        \includegraphics[width=0.47\textwidth]{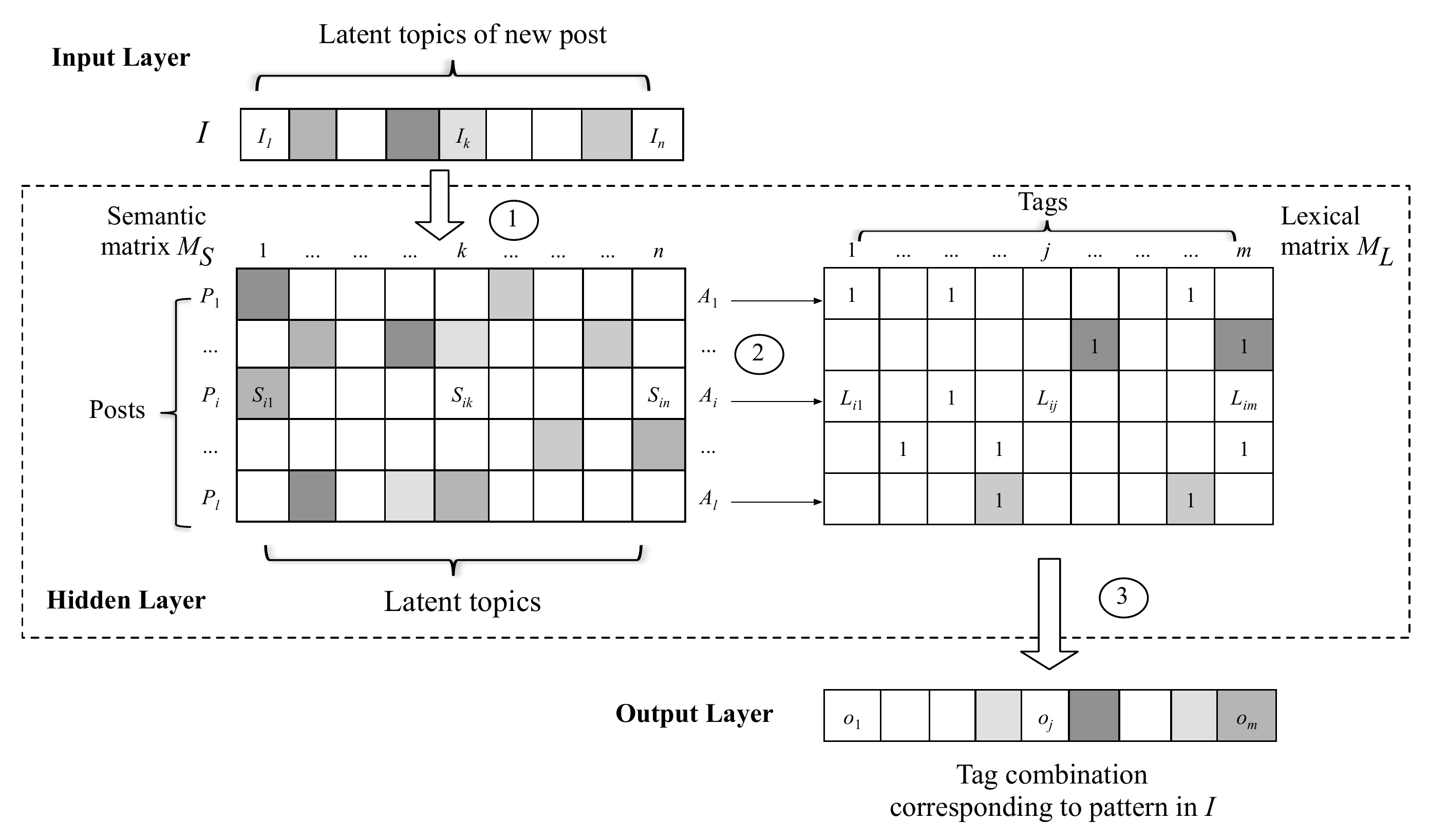}
        \caption{Schematic illustration of our basic 3Layers (3L) approach showing the connections between the semantic matrix ($M_S$) encoding the latent topics and the lexical matrix ($M_L$) encoding the tags.}				
	\label{fig:approach}
\end{figure}	

\subsection{Categorizing Resources} \label{sec:categorize}
The basic version of our proposed approach is solely based on human categorization processes \cite{kintsch2011construction}. It is termed 3Layers (3L) and is schematically represented in Figure \ref{fig:approach}. Similar to Kwantes \cite{kwantes2005using}, we apply a mechanism from MINERVA2, a computational theory of human categorization \cite{hintzman1984minerva}, to process the network constituted by the input, hidden and output layers shown in Figure \ref{fig:approach}. First, the latent semantic topics of the resource to be tagged are represented in the input layer in the form of vector $I$ with $n$ features (i.e., the latent topics). The latent semantic topics of the resources have been calculated in advance using Latent Dirichlet Allocation (LDA) \cite{krestel2009latent} based on the given tag distributions of the resources with a number of latent topics $Z$ of 1000 \cite{paul2013}.

Subsequently, $I$ is forwarded to the hidden layer, which represents the past posts of a user as a semantic matrix, $M_S$ ($l$ posts $\cdot$ $n$ latent topics matrix), and an interconnected lexical matrix, $M_L$ ($l$ posts $\cdot$ $m$ tags matrix). This way, each post of the user is represented by two associated vectors: a semantic vector of latent topics $S_{i,k}$ stored in $M_S$ and a verbatim vector of tags $L_{i,j}$ stored in $M_L$. $I$ acts as a cue to activate each post ($P_i$) in $M_S$ depending on the cosine-similarity ($Sim_i$) between both vectors, i.e., $I$ and $P_i$. To transform the resulting similarity values into activation values ($A_i$) and to reduce the influence of very low similarity values, $Sim_i$ is raised to the power of 3, i.e., $A_i = Sim_i^3$ (see \cite{kwantes2005using}).

Finally, these activation values are propagated to $M_L$ to activate tags that are associated with highly activated posts in the semantic matrix $M_S$ (circled numbers 2 and 3 in Figure \ref{fig:approach}). This step finalizes our basic 3L algorithm and is accomplished via the following equation that yields the value $o_j$ for each of the $m$ tags on the output layer:
\begin{align} \label{eq:3l}
	o_{j} = \underbrace{\sum_{i=1}^{l}(L_{i,j} \cdot A_{i})}_{3L}
\end{align}

\subsection{Temporal Dynamics} \label{sec:temporal}
To refine our approach drawing on temporal dynamics \cite{polyn2009context}, we integrate a time (or recency) component $T$ to assign higher activation values to tags that have been used more recently. As shown by Anderson \& Schooler \cite{anderson_reflections_1991}, the temporal decay of the users' word choices follows a power-law function. Thus, it can be modeled via the activation equation from the cognitive model ACT-R \cite{Anderson2004} (see also \cite{kowald2015refining}).

We use a simplified version of this equation to calculate the time component $T$: $BLL(j) = ln((t_{ref} - t_{j})^{-d})$, where $t_{ref}$ is the timestamp of the most recent post of the user and $t_{j}$ is the timestamp of the last occurrence of tag $j$ in the user's posts. The exponent $d$ accounts for the power-law of temporal decay of the user's tag choices and is typically set to $.5$ \cite{Anderson2004}. Summed up, 3LT, our time-dependent extension of 3L, can be realized using the following equation:
\begin{align} \label{eq:3lt}
	o^{T}_{j} = \underbrace{\sum_{i=1}^{l}(L_{i,j} \cdot BLL(j) \cdot A_{i})}_{3LT}
\end{align}

\subsection{Imitating Tags} \label{sec:imitate}
Research on social tagging \cite{floeck2011imitation,Seitlinger2012} has shown that a substantial variance in a user's tag choices can be explained by her tendency to imitate tags previously assigned by other users to a resource. Furthermore, modeling this imitation process allows recommending new tags, i.e., tags that were not used by the current user before. We realize tag imitation by taking into account the most popular tags in the tag assignments of the resource (i.e., MP$_{r}$ \cite{jaschke2007tag}). This approach was also chosen by other researchers in the field (e.g., \cite{zhang2012integrating,kowald2015refining}). Taken together, the list of $k$ recommended tags $RecTags(u, r)$ according to our proposed 3LT+MP$_r$ approach for the current user $u$ and the current resource $r$ is given by:
  \begin{align}
		RecTags(u, r) = \argmax_{j \in \text{Tags}}^{k}(\underbrace{\beta \|o^{T}_{j}\| + (1 - \beta) \||Y_{j, r}|\|}_{3LT+MP_r})
  \end{align}
where $|Y_{j, r}|$ is the number of assignments of tag $j$ for $r$ and $\beta$ can be used to inversely weigh the two components. $\beta$ was set to .5 to assign equal weights to individual and collective processes. Furthermore, we normalized the values of $o^{T}_{j}$ and $|Y_{j, r}|$ in order to combine them \cite{kowald2015refining}.

Our proposed approach presented in this section is fully implemented within our open-source and Java-based \textit{TagRec} framework \cite{domi2014a}, which is freely available via our Github Repository\footnote{\url{https://github.com/learning-layers/TagRec/}}. Among others, this framework also contains the baseline algorithms discussed in Section \ref{sec:relwork} and the evaluation method described in Section \ref{sec:eval}.

\section{Methodology} \label{sec:meth}
In this section we describe the methodology to validate our novel approach, including the datasets and evaluation method used.

\subsection{Datasets} \label{sec:datasets}
For reasons of reproducibility, we focused on three well-known folksonomy datasets, that are freely available for scientific purposes. Hence, we utilized datasets from the social tagging and publication sharing system BibSonomy\footnote{\url{http://www.kde.cs.uni-kassel.de/bibsonomy/dumps/}}, the reference management system CiteULike\footnote{\url{http://www.citeulike.org/faq/data.adp}} and the social bookmarking platform Delicious\footnote{\url{https://www.uni-koblenz.de/FB4/Institutes/IFI/AGStaab/Research/DataSets/PINTSExperimentsDataSets/}}.

We excluded all automatically-generated tags from the datasets (e.g., \textit{no-tag} or \textit{bibtex-import}) and decapitalized all tags as suggested in related work (e.g., \cite{Rendle2010}). Furthermore, to reduce the computational effort, we randomly selected 10\% of CiteULike and 3\% of Delicious user profiles (see also \cite{gemmell2009improving}) but did not apply a $p$-core pruning to avoid a biased evaluation (see \cite{Doerfel2013}). The statistics of our used dataset samples are shown in Table \ref{tab:datasets}.

\begin{table}[t!]
  \setlength{\tabcolsep}{2.6pt}	
  \centering
    \begin{tabular}{l|lllll}
    \specialrule{.2em}{.1em}{.1em}
      Dataset			& $|P|$			& $|U|$	& $|R|$	& $|T|$	& $|TAS|$	 							\\\hline 
      BibSonomy	  & 400,983 & 5,488  & 346,444		& 103,503	& 1,479,970				\\\hline
			CiteULike		& 379,068 & 8,322  & 352,343		& 138,091	& 1,751,347				\\\hline
			Delicious		& 1,416,151 & 15,980  & 931,993		& 180,084	& 4,107,107			\\			
		\specialrule{.2em}{.1em}{.1em}								
    \end{tabular}
    \caption{Properties of the datasets, where $|P|$ is the number of posts, $|U|$ is the number of users, $|R|$ is the number of resources, $|T|$ is the number of tags and $|TAS|$ is the number of tag assignments.}	
  \label{tab:datasets}
\end{table}

\begin{figure*}[t!]
   \centering 
   \subfloat[BibSonomy]{ 
      \includegraphics[width=0.324\textwidth]{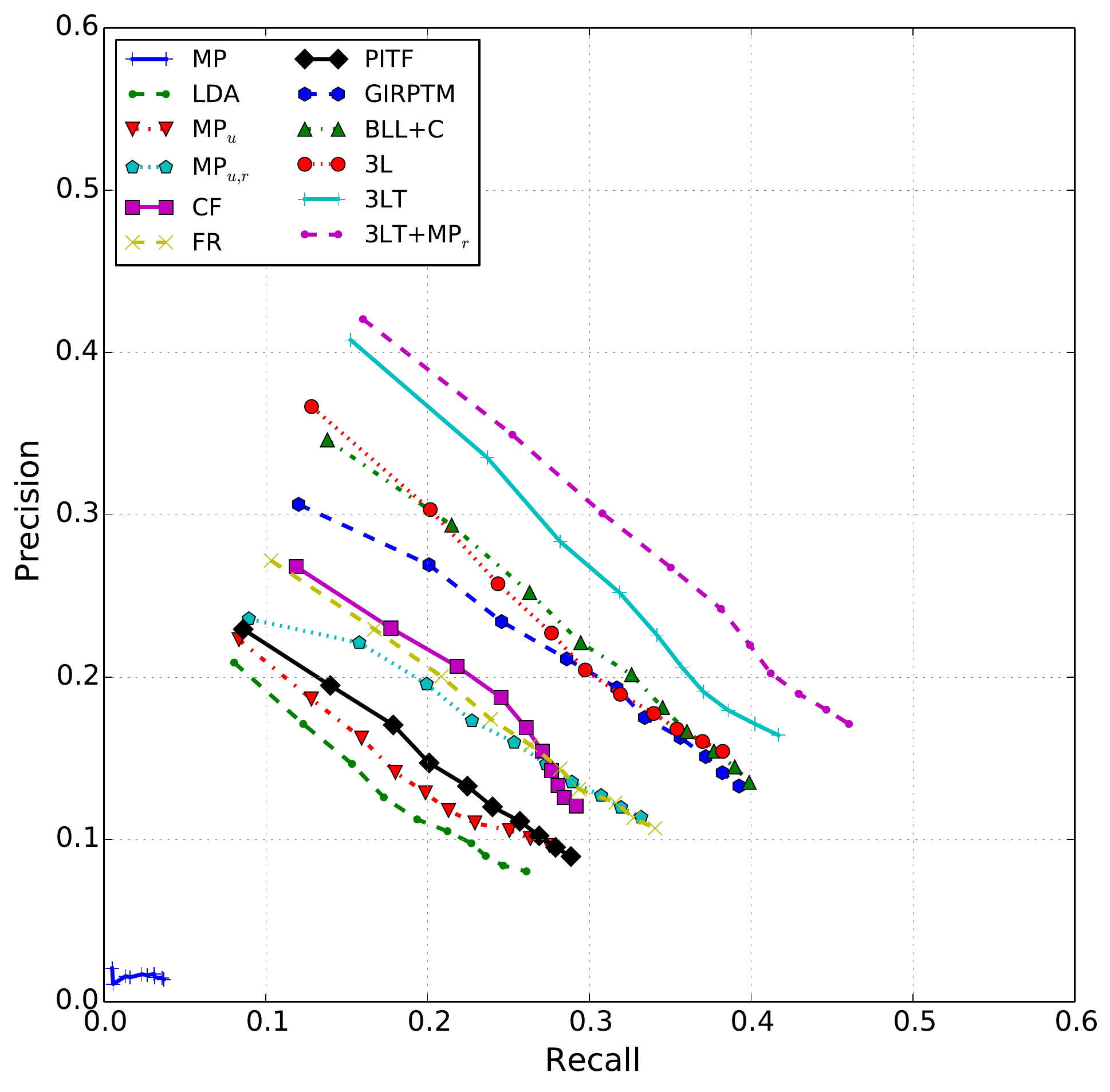} 
   } 
   \subfloat[CiteULike]{ 
      \includegraphics[width=0.324\textwidth]{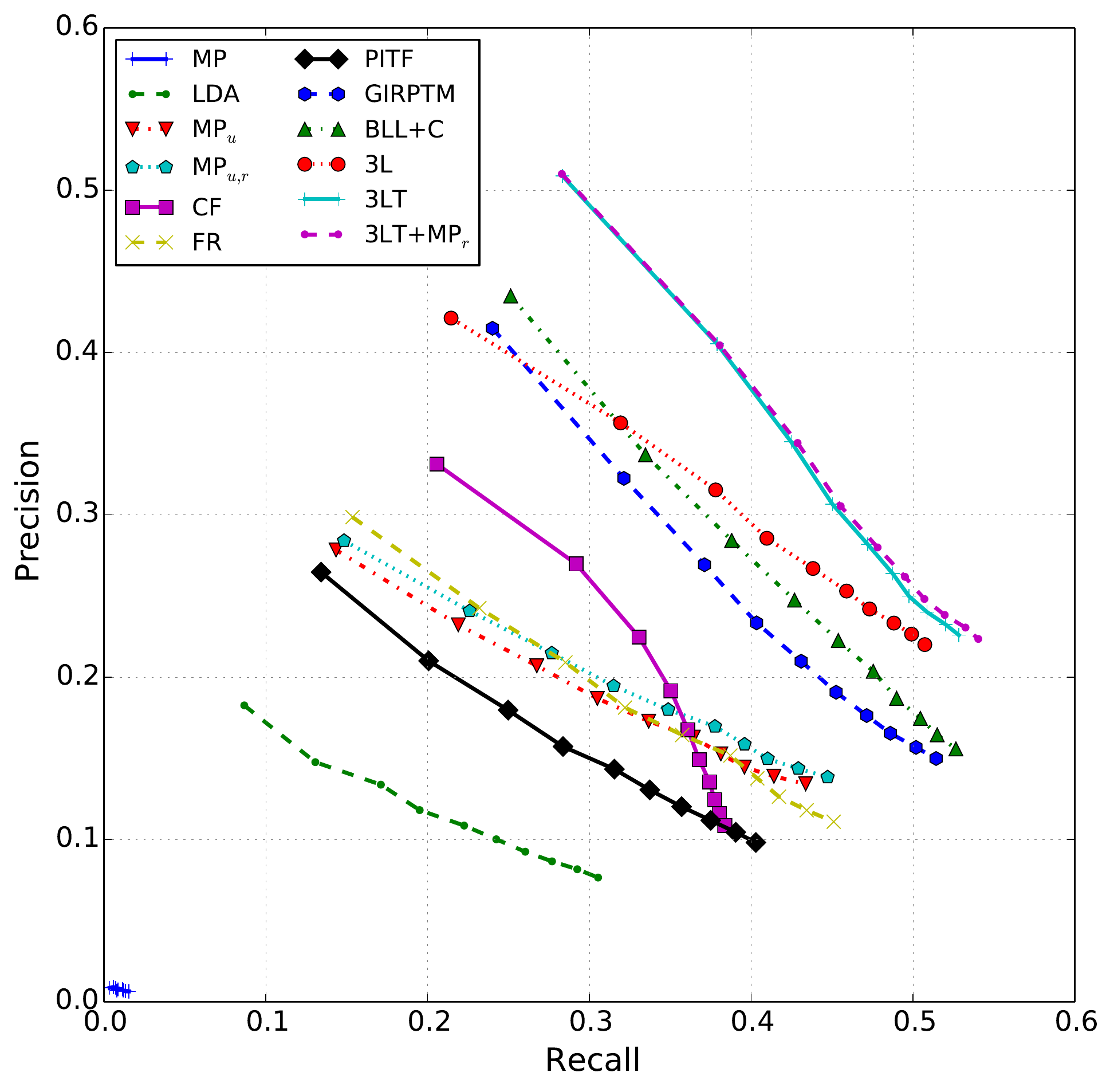} 
   } 
   \subfloat[Delicious]{ 
      \includegraphics[width=0.324\textwidth]{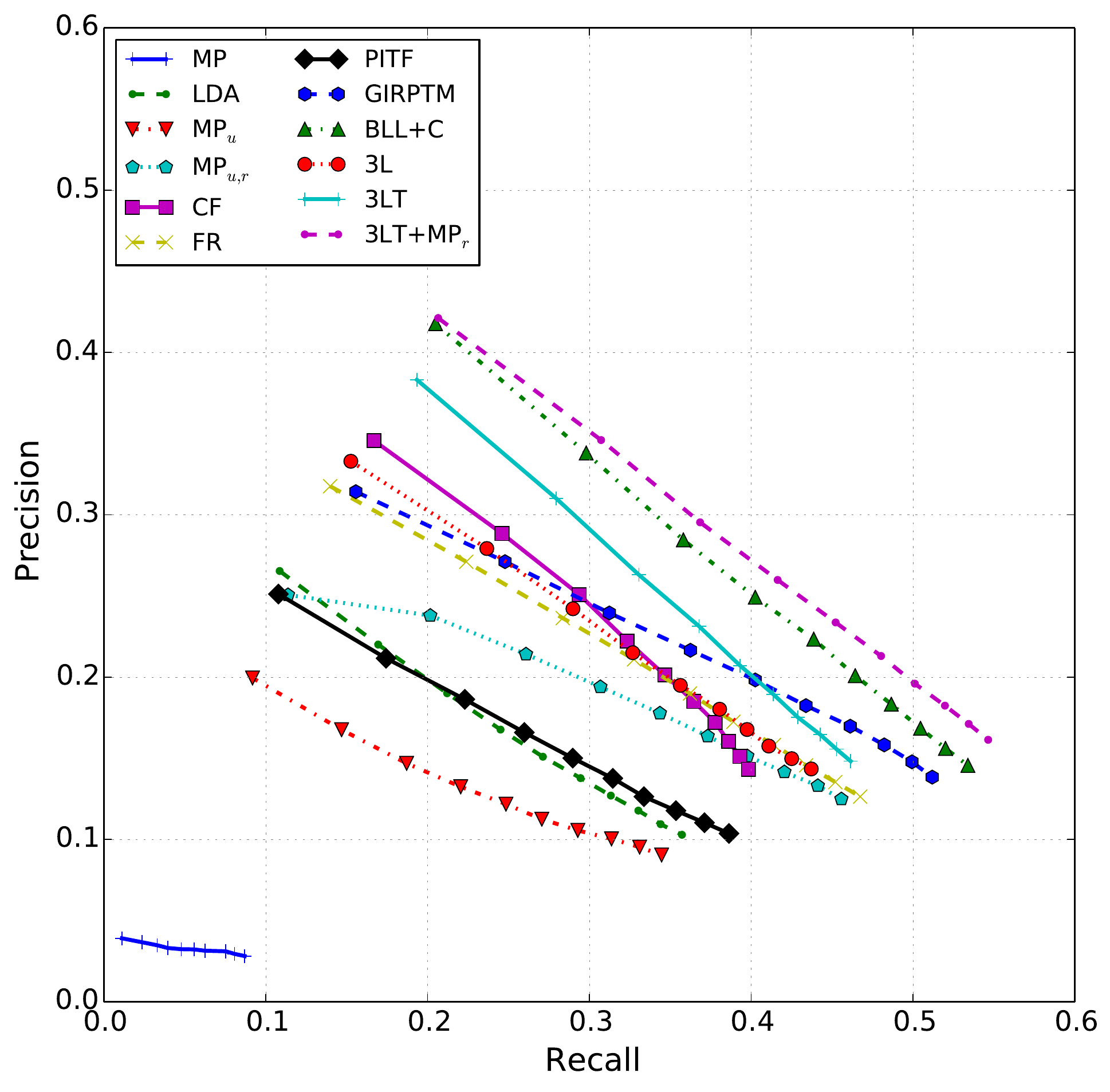} 
   }
   \caption{Precision/Recall plots for BibSonomy, CiteULike and Delicious showing the performance of our approach (3LT+MP$_r$) compared to 3L, 3LT and state-of-the-art algorithms for $k$ = 1 - 10 recommended tags.}
	 \label{fig:prec_rec}
\end{figure*}

\subsection{Evaluation Method} \label{sec:eval}
To evaluate our tag recommender approach, we followed a standard procedure in recommender research (e.g., \cite{jaschke2007tag}) and split the three datasets into training and test sets. In order to preserve the chronological order of the data, for each user we selected her most recent post (in time) and placed it into the test set. The remaining posts were then used to train the algorithms. This procedure is a promising simulation of a real-world environment, since it predicts the user's future tagging behavior based on her tagging behavior in the past \cite{campos2013time}. To ensure that a minimum amount of tagging ``history'' is available for training, we focused on users with at least 20 posts. 
We conducted this evaluation by applying a post-filtering method: while recommendations were calculated based on the entire folksonomy graph, accuracy estimates were computed only on the basis of the filtered user profiles. This resulted in 780 users in the case of BibSonomy, 1,757 in the case of CiteULike and 7,469 in the case of Delicious.

In order to finally quantify the recommender quality and to benchmark our approach against other tag recommendation algorithms mentioned in Section \ref{sec:relwork}, we compared the top-$10$ tags an algorithm suggested for a user-resource pair with the set of relevant tags in the corresponding post in the test set. Based on these comparisons, various evaluation metrics can be calculated, that have originated from information retrieval and recommender systems research (e.g., Precision, Recall, F1-score, MRR, MAP, nDCG) \cite{jaschke2007tag,lipczak2012hybrid}. At the moment, in this work we focus on Precision and Recall for $k$ = 1 - 10 recommended tags.

\section{Preliminary Results} \label{sec:results}
The preliminary results of our evaluation for BibSonomy, CiteULike and Delicious are shown in the three Precision/ Recall plots in Figure \ref{fig:prec_rec}. When looking at the results for the baseline algorithms (see Section \ref{sec:relwork}), it is apparent that, as expected, all personalized algorithms outperform the unpersonalized MP approach. Additionally, the highest accuracy estimates among these baselines are reached by the two time-dependent methods GIRPTM and BLL+C. This emphasizes the importance of taking temporal dynamics into account in tag recommender research. Moreover, BLL+C, which uses a power-law decay function, outperforms GIRPTM, which relies on an exponential decay function.

A comparison of the results for the basic version of our proposed approach 3L, which is solely based on human categorization processes, with 3LT, which further integrates temporal dynamics, shows that, as expected, 3LT provides higher Recall and Precision values than 3L. This further proves that temporal dynamics play an important role in tag prediction tasks. Additionally, the complete version of our approach 3LT+MP$_r$, which also utilizes macro-level processes in the form of imitating the most popular tags assigned to the target resource by other users (MP$_r$), provides better results than 3LT. This is especially true in the case of Delicious, where numerous tags from other users are available for imitation.

Apart from that, and even more important, our proposed approach 3LT+MP$_r$ also outperforms the current state-of-the-art algorithms mentioned in Section \ref{sec:relwork} in all three settings. This includes the time-based GIRPTM and BLL+C approaches, that also utilize temporal dynamics and tag imitation but ignore categorization, which further reveals the importance of all three examined types of cognitive processes. In brief, the results of our experiments not only show that an interplay between individual micro-level and collective macro-level processes can be used to develop an effective tag recommender approach but also that such an approach can outperform current state-of-the-art algorithms.

\section{Conclusion and Future Work} \label{sec:conclusion}
This thesis aims at modeling the cognitive micro- and macro-level processes that play a role when people assign tags to resources. At the current stage of this thesis, this is achieved by the development and evaluation of a novel tag recommender termed 3LT+MP$_r$. The preliminary results of our evaluation for datasets gathered from BibSonomy, CiteULike and Delicious show that 3LT+MP$_r$ can not only outperform current state-of-the-art algorithms but also provides higher accuracy estimates than 3L and 3LT that implement only some of the examined processes. These results corroborate our hypothesis from Section \ref{sec:problems}, that our theory-driven tag recommender can not only improve recommender accuracy in general but also can help to better understand the underlying cognitive processes. Additionally, we introduce an open-source tag recommender benchmarking framework termed TagRec (see also \cite{domi2014a}), which contains not only our proposed approach but also standardized baseline algorithms and evaluation methods.

At present, one limitation of this thesis is that we only focus on one cognitive model, namely MINERVA2, to account for the individual micro-level processes in social tagging systems. With that regard, in the future we would like to use and evaluate also another cognitive model (e.g., ACT-R) to realize these (and also additional) processes. We also plan to further improve our approach by investigating content data of the resources (e.g., title and description), which could highly expand the set of possible tags that can be recommended. Moreover, we will evaluate our proposed approach in terms of not only recommender accuracy but also computational costs in order to validate the efficiency of our theory-driven approach. We also plan to integrate our approach into a real online social tagging system, since only then it will be possible to examine our tag recommender's performance with regard to user acceptance. Finally, we would like to adapt the idea of our approach also for other problems in the recommender systems domain, such as the recommendation of resources, topics and users.

\textbf{Acknowledgments:}
The author would like to thank Elisabeth Lex, Tobias Ley, Paul Seitlinger and Christoph Trattner for valuable discussions related to this thesis. This work is funded by the Know-Center and the EU-funded project Learning Layers (Grant Agreement 318209). The Know-Center is funded within the Austrian COMET Program - Competence Centers for Excellent Technologies - under the auspices of the Austrian Ministry of Transport, Innovation and Technology, the Austrian Ministry of Economics and Labor and by the State of Styria. COMET is managed by the Austrian Research Promotion Agency (FFG).

\bibliographystyle{abbrv}

\begin{thebibliography}{10}

\bibitem{Anderson2004}
J.~R. Anderson, M.~D. Byrne, S.~Douglass, C.~Lebiere, and Y.~Qin.
\newblock An integrated theory of the mind.
\newblock {\em Psychological Review}, 111(4):1036--1050, 2004.

\bibitem{anderson_reflections_1991}
J.~R. Anderson and L.~J. Schooler.
\newblock Reflections of the environment in memory.
\newblock {\em Psychological Science}, 2(6):396--408, 1991.

\bibitem{campos2013time}
P.~G. Campos, F.~D{\'\i}ez, and I.~Cantador.
\newblock Time-aware recommender systems: a comprehensive survey and analysis
  of existing evaluation protocols.
\newblock {\em User Modeling and User-Adapted Interaction}, pages 1--53, 2013.

\bibitem{Dellschaft2012}
K.~Dellschaft and S.~Staab.
\newblock Measuring the influence of tag recommenders on the indexing quality
  in tagging systems.
\newblock In {\em Proc. of HT'12}, pages 73--82, New York, NY, USA, 2012. ACM.

\bibitem{Doerfel2013}
S.~Doerfel and R.~J\"{a}schke.
\newblock An analysis of tag-recommender evaluation procedures.
\newblock In {\em Proc. of Recsys'13}, pages 343--346, New York, NY, USA, 2013.
  ACM.

\bibitem{floeck2011imitation}
F.~Floeck, J.~Putzke, S.~Steinfels, K.~Fischbach, and D.~Schoder.
\newblock Imitation and quality of tags in social bookmarking
  systems--collective intelligence leading to folksonomies.
\newblock In {\em On collective intelligence}, pages 75--91. Springer, 2011.

\bibitem{fu2008microstructures}
W.-T. Fu.
\newblock The microstructures of social tagging: a rational model.
\newblock In {\em Proc. of CSCW'08}, pages 229--238. ACM, 2008.

\bibitem{gemmell2009improving}
J.~Gemmell, T.~Schimoler, M.~Ramezani, L.~Christiansen, and B.~Mobasher.
\newblock Improving folkrank with item-based collaborative filtering.
\newblock {\em Recommender Systems \& the Social Web}, 2009.

\bibitem{hintzman1984minerva}
D.~L. Hintzman.
\newblock Minerva 2: A simulation model of human memory.
\newblock {\em Behavior Research Methods, Instruments, \& Computers},
  16(2):96--101, 1984.

\bibitem{hotho2006information}
A.~Hotho, R.~J{\"a}schke, C.~Schmitz, and G.~Stumme.
\newblock Information retrieval in folksonomies: Search and ranking.
\newblock In {\em The semantic web: research and applications}, pages 411--426.
  Springer, 2006.

\bibitem{jaschke2007tag}
R.~J{\"a}schke, L.~Marinho, A.~Hotho, L.~Schmidt-Thieme, and G.~Stumme.
\newblock Tag recommendations in folksonomies.
\newblock In {\em Proc. of PKDD'07}, pages 506--514. Springer, 2007.

\bibitem{jaschke2008tag}
R.~J{\"a}schke, L.~Marinho, A.~Hotho, L.~Schmidt-Thieme, and G.~Stumme.
\newblock Tag recommendations in social bookmarking systems.
\newblock {\em Ai Communications}, 21(4):231--247, 2008.

\bibitem{kintsch2011construction}
W.~Kintsch and P.~Mangalath.
\newblock The construction of meaning.
\newblock {\em Topics in Cognitive Science}, 3(2):346--370, 2011.

\bibitem{Korner2010}
C.~K\"{o}rner, D.~Benz, A.~Hotho, M.~Strohmaier, and G.~Stumme.
\newblock Stop thinking, start tagging: tag semantics emerge from collaborative
  verbosity.
\newblock In {\em Proc. of WWW'10}, WWW '10, pages 521--530, New York, NY, USA,
  2010. ACM.

\bibitem{kowald2015refining}
D.~Kowald, S.~Kopeinik, P.~Seitlinger, T.~Ley, D.~Albert, and C.~Trattner.
\newblock Refining frequency-based tag reuse predictions by means of time and
  semantic context.
\newblock In {\em Mining, Modeling, and Recommending Things' in Social Media},
  pages 55--74. Springer, 2015.

\bibitem{domi2014a}
D.~Kowald, E.~Lacic, and C.~Trattner.
\newblock Tagrec: Towards a standardized tag recommender benchmarking
  framework.
\newblock In {\em Proc. of HT'14}, New York, NY, USA, 2014. ACM.

\bibitem{krestel2009latent}
R.~Krestel, P.~Fankhauser, and W.~Nejdl.
\newblock Latent dirichlet allocation for tag recommendation.
\newblock In {\em Proc. of Recsys'09}, pages 61--68. ACM, 2009.

\bibitem{kwantes2005using}
P.~J. Kwantes.
\newblock Using context to build semantics.
\newblock {\em Psychonomic Bulletin \& Review}, 12(4):703--710, 2005.

\bibitem{lipczak2012hybrid}
M.~Lipczak.
\newblock {\em Hybrid Tag Recommendation in Collaborative Tagging Systems}.
\newblock PhD thesis, Dalhousie University, 2012.

\bibitem{marinho2008collaborative}
L.~B. Marinho and L.~Schmidt-Thieme.
\newblock Collaborative tag recommendations.
\newblock In {\em Data Analysis, Machine Learning and Applications}, pages
  533--540. Springer, 2008.

\bibitem{polyn2009context}
S.~M. Polyn, K.~A. Norman, and M.~J. Kahana.
\newblock A context maintenance and retrieval model of organizational processes
  in free recall.
\newblock {\em Psychological review}, 116(1):129, 2009.

\bibitem{rendle2010factorization}
S.~Rendle.
\newblock Factorization machines.
\newblock In {\em Proc. of ICDM'10}, pages 995--1000. IEEE, 2010.

\bibitem{Rendle2010}
S.~Rendle and L.~Schmidt-Thieme.
\newblock Pairwise interaction tensor factorization for personalized tag
  recommendation.
\newblock In {\em Proc. of WSDM'10}, pages 81--90, New York, NY, USA, 2010.
  ACM.

\bibitem{paul2013}
P.~Seitlinger, D.~Kowald, C.~Trattner, and T.~Ley.
\newblock Recommending tags with a model of human categorization.
\newblock In {\em Proc. of CIKM'13}, pages 2381--2386, New York, NY, USA, 2013.
  ACM.

\bibitem{Seitlinger2012}
P.~Seitlinger and T.~Ley.
\newblock Implicit imitation in social tagging: familiarity and semantic
  reconstruction.
\newblock In {\em Proc. of CHI'12}, pages 1631--1640, New York, NY, USA, 2012.
  ACM.

\bibitem{zhang2012integrating}
L.~Zhang, J.~Tang, and M.~Zhang.
\newblock Integrating temporal usage pattern into personalized tag prediction.
\newblock In {\em Web Technologies and Applications}, pages 354--365. Springer,
  2012.

\end{thebibliography}

\end{document}